\documentclass[aps,twocolumn,groupedaddress,showpacs,nofootinbib]{revtex4}

\usepackage{graphicx}% Include figure files
\usepackage{hyperref}
\usepackage{dcolumn}% Align table columns on decimal point
\usepackage{bm}% bold math
\usepackage{epsfig,amsbsy,bm,amssymb,amsmath}
\usepackage{xcolor} 
\usepackage{tikz}
\usetikzlibrary{shadings}

\renewcommand{\k}{{k}}
\newcommand{\hs}{\hspace*}

\newcommand{\w}{\omega}

\newcommand{\eref}[1] {(\ref{#1})}
\newcommand{\Eref}[1] {Eq.~(\ref{#1})}
\newcommand{\Fref}[1] {Fig. \ref{#1}}
\newcommand{\Tref}[1] {Table \ref{#1}}

\newcommand{\be}{\begin{equation}}
\newcommand{\ee}{\end{equation}}
\newcommand{\br}{\begin{eqnarray*}}
\newcommand{\er}{\end{eqnarray*}}
\newcommand{\ba}{\begin{eqnarray}}
\newcommand{\ea}{\end{eqnarray}}
\newcommand{\bp}{\begin{minipage}}
\newcommand{\ep}{\end{minipage}}
\newcommand{\bt}{\begin{tabular}}
\newcommand{\et}{\end{tabular}}

\newcommand{\bs}{\bigskip}
\newcommand{\ms}{\vspace*{-5mm}}
\newcommand{\mms}{\vspace*{-2.5mm}}

\renewcommand{\r}{{\bm r}}
\renewcommand{\k}{{\bm k}}

   \newcommand{\E}{{\bm E}}

  \newcommand{\mc}{\multicolumn}
 \newcommand{\ig}[1]{\includegraphics[width={#1}]}

\newcommand{\rain}{$r$RABBITT~}
\newcommand{\crain}{$cr$RABBITT~}
\newcommand{\Sref}[1] {Sec.~\ref{#1}}

\renewcommand{\t}{\tau}

\renewcommand{\H}{H$_2$~}

\pgfdeclareverticalshading{rainbow}{200bp}
{color(0bp)=(red); color(25bp)=(red); color(35bp)=(yellow);
color(45bp)=(green); color(55bp)=(cyan); color(65bp)=(blue);
color(75bp)=(violet); color(100bp)=(violet)}

\newcommand{\Wcm}[2]{
$\rm {#1}\times10^{{#2}}~W/cm^2$}

\newcommand{\ask}[1]{\textcolor{blue}{#1}\hs{-1mm} }
\renewcommand{\ask}[1]{\textcolor{black}{#1}\hs{-1mm} }

\begin{document}
\bibliographystyle{apsrev}

\title{Continuous rainbow RABBITT investigation of resonant states in
  He and \H}

\author{Vladislav ~V. Serov$^{1}$}
\author{Anatoli~S. Kheifets$^{2}$}

%VS
\affiliation{$^{1}$Department of Medical Physics, Saratov State University, Saratov
  410012, Russia}
\email{vladislav.serov@gmail.com}

\affiliation{$^{2}$Research School of Physics, The Australian
  National University, Canberra ACT 2601, Australia}
\email{A.Kheifets@anu.edu.au}

 \date{\today}

\pacs{32.80.Rm 32.80.Fb 42.50.Hz}

\begin{abstract}
We employ Reconstruction of Attosecond Beating By Interference of
Two-photon Transitions with an advanced energy resolution (rainbow
RABBITT) to resolve under-threshold discrete excitations and
above-threshold auto\-ionizing states in the He atom and the \H
molecule. Both below and above the threshold, the whole series of
resonances is reconstructed continuously \ask{and at once} by the
parity based separation of the two-photon ionization amplitude. This
allows for an efficient extraction of the RABBITT magnitude and phase
parameters without the need for adjusting the laser photon
frequency. The latter parameters are then used to test the validity of
the logarithmic Hilbert transform which relates the RABBITT phase and
magnitude in the resonant region.
\end{abstract}

\maketitle
%\end{document}%STOP!!!

\ms\ms\ms
\ask{\section{Introduction}}
\ms

\ask{Reconstruction of Attosecond Beating by Interference of
  Two-photon Transitions (RABBITT) is a well established two-photon
  interferometric technique. In this technique, like in the pioneering
  work \cite{KlunderPRL2011}, an atomic target is ionized with a comb
  of odd XUV harmonics $(2q\pm1)\w$ from an attosecond pulse train
  (APT) and probed with a dressing IR laser field with the carrier
  frequency $\w$. The primary ionization results in regularly spaced
  harmonic peaks in the photoelectron energy spectrum. The dressing
  field adds side\-bands (SBs) centered at $2q\w$. The height of the
  sidebands oscillates with XUV/IR delay $\tau$ at twice the IR photon
  frequency
\be
S_{2q}(\tau) =A+
B\cos[2\omega\tau+C]
\ \ , \ \
C = 2\w\t_a
\ .
\label{RABBITT}
\ee
Here $A,B>0$ are the RABBITT magnitudes parameters and $C$ is its phase
which is related with the atomic time delay $\t_a$.}

\ask{The conventional RABBITT analyzes the whole SB as a single entity
  and determines the parameters in \Eref{RABBITT} as a function of the
  SB order $2q$.  The rainbow RABBITT, or $r$RABBITT, goes one step
  further. It looks inside the SB and determines the parameters in
  \Eref{RABBITT} as functions of the photo\-electron energy across the
  given SB.  Experimentally, such a detailed energy-resolved analysis
  of SB's is achieved by an enhanced spectral resolution which can
  bring a wealth of additional information.  In particular,} it can
help to disentangle various ionization pathways involving
auto\-ionizing resonances
\cite{Kotur2016,Gruson734,Busto2018,Isinger2019}, below threshold
discrete states \cite{Neoricic2022} and fine-structure splittings
\cite{Turconi2020,Roantree2023} The same technique can be beneficial
when the presence of multiple ionization channels leads to spectral
congestion in atoms \cite{Alexandridi2021} and molecules
\cite{Borras2023}.  Similarly to \rain, an advanced spectral
resolution of SB's can be achieved \ask{in conventional RABBITT} by
fine tuning of the laser photon frequency
\cite{SwobodaPRL2010,Barreau2019,Drescher2022}.

\ask{Simultaneous energy resolution of several SB's in \rain is not
  possible at present neither experimentally nor theoretically. Each
  SB should be resolved individually by adjusting the laser carrier
  frequency $\w$ appropriately. In the present theoretical work, we
  lift this restriction and overcome such an obstacle. We demonstrate
  that all the SB's in the photo\-electron spectrum can be resolved
  simultaneously and at once.}

Our simulations are based on a numerical solution of the
time-dependent Schr\"odinger equation (TDSE) driven by a combination
of the ionizing XUV and dressing IR laser fields. The ionizing field
is generated by an APT composed of a number of short XUV pulses. A
weak dressing field is represented by a femto\-second IR pulse.
By performing several TDSE calculations at varying $\omega$ at
sufficiently broad, but not yet overlapping primary and secondary
peaks, it becomes possible to obtain the RABBITT parameters
continuously across the photo\-electron spectrum of interest. It is
this strategy that is commonly employed in \rain experiments e.g. in
\cite{Busto2018,Neoricic2022}.

Menawhile, unlike in the experiment, the TDSE simulations give us
access to the ionization amplitudes. The primary harmonic peaks in the
photo\-electron spectrum arise from the constructive interference of
the one-photon ionization amplitudes which interfere destructively in
the gaps between these peaks. These gaps are filled by the SB's which
therefore originate solely from the two-photon ionization
amplitudes. The latter amplitudes can be obtained by subtracting the
single-photon XUV-only ionization amplitude from the combined XUV+IR
amplitude. Such a subtraction is not possible experimentally as the
two-photon ionization cross-section would contain an XUV/IR
interference term. For the amplitude subtraction, it is suffice to
use a sufficiently short XUV pulse with a broad spectrum
\cite{PhysRevA.93.063417,Serov2017}.  The probability of the
two-photon ionization process will be the function of the phase shift
between the XUV and IR pulses just as prescribed by \Eref{RABBITT}. In
this way, it is possible to obtain the RABBITT parameters continuously
across the whole photo\-electron spectrum.  We refer to such an
approach as a continuous \rain or \crain.

In the case of an atom or a symmetric molecule, instead of
subtraction, the two-photon process amplitude can be isolated based on
its parity. If the parity of the initial state is positive, the parity
of the one-photon absorption amplitude will be negative, while the
two-photon process amplitude will have positive parity.  This approach
requires even fewer calculations.

\ask{We demonstrate the advantages of the \crain technique by
  studying} formation of auto\-ionizing states (AIS) in the He atom
and the \H molecule. In addition, we look under the threshold and map
a series of discrete one-electron excitation\-s. Such under-threshold
or $u$RABBITT investigations showed their great potential in mapping
the target atom electronic structure
\cite{SwobodaPRL2010,Villeneuve2017,PhysRevA.103.L011101,Kheifets2021Atoms,Neoricic2022,Kheifets2023,moioli2024role}

After the \crain phase and amplitude parameters are obtained across
the whole photo\-electron spectrum, they can be used to test the
validity of the Kramers-Kronig relation in the form of the logarithmic
Hilbert transform (LHT). This relation connects the photo\-ionization
phase and amplitude in the resonant region
\cite{Ji2024,kheifets2024resonant}. So far this relation has been
tested for single-photon ionization. The present work demonstrates the
first successful application of the LHT to two-photon resonant
ionization processes.

\ask{The rest of the paper is organized as follows. In
  \Sref{Technique} we describe our computational technique. Our
  numerical results are presented in \Sref{Results}. The latter
  section is divided into \Sref{AIS} and \Sref{Under} dealing with
  auto\-ionizing states and below-threshold excitation\-s,
  respectively. We conclude in \Sref{Conclusion} by summarizing our
  main findings and outlining possible extension of the present work.}

\bs
\section{Computational technique}
\label{Technique}  
As in our previous applications \cite{Serov2024,serov2024fano}, we
solve numerically the two-electron TDSE
\be
i(\partial/\partial t) \Psi(\r_1, \r_2,t) =
\hat{H} \Psi(\r_1, \r_2,t) 
\ ,
\label{TDSE}
\ee 
where the Hamiltonian $\hat{H}$ contains the single-electron ionic
parts $\hat{h}_0(\r)$, the electron-electron interaction
$v(\r_1,\r_2)$ and the interaction with an external field
$\hat{w}(\r,t)=- \E(t) \cdot \r $ expressed in the coordinate gauge.

Solution of the TDSE \eref{TDSE} is sought as 
a multi-configuration expansion
\be
\Psi(\r_1, \r_2,t) = \sum_{n=1}^{N_s}
    [\psi_n(\r_1,t) \phi_n(\r_2) + (-1)^S
      \phi_n(\r_1) \psi_n(\r_2,t) ] ,
\label{multi}
\ee
built from the ionic eigen\-states satisfying the stationary
Schr\"odinger equation
$
\hat{h}_0 \phi_n(\r) = \epsilon_n \phi_n(\r) \ .
$
Further details on construction of the multi\-configuration basis and
its symmetrization are given in \cite{Serov2024}. Numerical details
that are pertinent to the TDSE solution can be found in 
\cite{PhysRevA.84.062701,Puzynin2000,PhysRevA.88.043403}.

The TDSE \eref{TDSE} is driven by the electric field $E(t)$ composed
of co-linearly $\hat z$-polarized XUV and IR fields.
The XUV field is represented  by an APT
\be
E_{\rm APT}(t)=\sum_{\nu=-\left\lfloor N_{\rm
APT}/2\right\rfloor}^{\left\lfloor N_{\rm APT}/2\right\rfloor}(-1)^\nu
f_{\rm env}(t_\nu) E_{\rm XUV}(t-t_\nu) 
 \ ,
\label{A_APT}
\ee
where $N_{\rm APT}=41$ and the arrival time of each pulse
$
t_\nu=\nu T_{\rm IR}/2
$
is a half integer of the period of the IR oscillation $T_{\rm  IR}=2\pi/\omega$.
The envelope of the APT is modeled as
\be
f_{\rm env}(t)=\exp\left[-2\ln 2 (t/\tau_{\rm APT})^2\right]
\ ,
\ee
where $\tau_{\rm APT}$ is the full width at half maximum (FWHM) of
the train.
Each ultra\-short XUV pulse in the train takes the form
\be E_{\rm XUV}(t)=E_{\rm XUV}\exp\left[-2\ln 2 (t/\tau_{\rm
    XUV})^2\right]\cos\Omega t \ , \ee
with   $\tau_{\rm XUV}$ being the FWHM of the  pulse. 
The XUV central frequency  $\Omega$ and the time constants
$\tau_{\rm XUV}, \tau_{\rm APT}$ are chosen to span a sufficient
number of harmonics in the range of photon frequencies of interest for
a given target.
The IR pulse has a cos$^4$ envelope and is shifted from the center of the
APT by a variable delay $\tau$. The XUV and IR intensities are
\Wcm{1}{14} and \Wcm{1}{10}, respectively.
As both the XUV and IR fields are weak, we can calculate contributions of
each XUV pulse to ionization, and after that summarize this
contributions to get ionization amplitude for all the train
\cite{PhysRevA.93.063417}. Due to such a split, we can assume evolution
 in the field of just a single XUV pulse dressed by the IR  field
\be
E_\nu(t)= E_{\rm XUV}(t) + E_{\rm IR}(t-\tau-t_\nu). \label{amp1XUV+IR}
\ee
The resulting ionization amplitude $A_{n\nu}(\k)$ in such a field
becomes a function of the photo\-electron momentum $\k$ and the
quantum number $n$ characterizing  the residual ion. It can be used to
construct the two-photon XUV+IR ionization amplitude in the XUV+IR
field by performing the summation \cite{PhysRevA.93.063417}
\be
A_n(\k)=\hs{-5mm}\sum_{\nu=-\left\lfloor N_{\rm
    APT}/2\right\rfloor}^{\left\lfloor N_{\rm APT}/2\right\rfloor}\hs{-5mm}
(-1)^\nu f_{\rm env}(t_\nu)
e^{i\left(\epsilon_n+k^2/2-E_0\right)t_\nu}
A_{n\nu}(\k) 
\ ,
\label{sum_amplRABBIT}
\ee
where $E_0$ is the ground state energy.  
If the IR pulse is so long that the change in its envelope over time
$\tau_{\rm APT}$ is negligible, it is sufficient to calculate
$A_{n0}(\k)$ and $A_{n1}(\k)$ only, because in this case
$A_{n\nu}(\k)\simeq A_{n,|\nu|\, {\rm mod}\, 2}(\k)$.
The pure two-photon ionization amplitude can be expressed as the
difference 
 $A_{n\,{\rm 2photon}}(\k)=A_{n0}(\k)-A_{n\,{\rm
    XUV}}(\k)\ .$

If the system under consideration is an atom or a symmetric molecule,
the number of required calculations is further reduced, as in this
case, $A_{n1}(\k) = P_n A_{n0}(-\k)$, where $P_n$ is the parity of the
$n$'s ionic state.  The pure two-photon ionization amplitude can be
extracted by a simple symmetrization
$A_{n}(\k) = [ A_{n0}(\k) + P_n
  A_{n0}(-\k) ] /2 \ .$

\ask{
When solving the TDSE \eref{TDSE}, the radial dependence of the wave
function is handled by the discrete variable representation (DVR) on
the finite elements of the Gauss-Lobatto quadrature. The angular
dependence was represented by a spherical harmonic expansion
\cite{PhysRevA.84.062701}.
The time evolution in \Eref{TDSE} was carried out using an implicit
fourth-order scheme \cite{Puzynin2000}, which is a generalization of
the Crank-Nicholson scheme. The system of linear equations to which
this scheme leads was solved using the bi\-conjugate gradient method
with a pre\-conditioner.
Suppression of the unphysical reflection from the radial grid
boundary was carried out using exterior complex scaling (ECS) for the
radial variable.
Ionization amplitudes were extracted from the calculated wave function
using the t-SURFFc method \cite{PhysRevA.88.043403}.}

\ms
\section{Results}
\label{Results}

\subsection{Auto\-ionizing states}
\label{AIS}

\begin{figure}[t]
\ig{8cm}{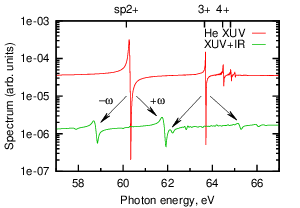}%{Data/Fig1.eps}
\caption{Photoelectron spectra of He in single XUV photon (top, red) and
  XUV+IR  (bottom, green) ionization. The corresponding AIS are marked
on the top horizontal axis according to \cite{Domke1996}. The arrows
indicate the genesis of the resonant peaks in the XUV+IR spectrum. The
laser photon energy $\w=1.53$~eV}
\label{Fig1}
\end{figure}

The continuously resolved spectrum of the AIS in He is displayed in
\Fref{Fig1}. The AIS energies and identification are marked on the top
horizontal axis according to \cite{Domke1996}. The top (red) spectrum
corresponds to XUV ionization only. Such a spectrum can be compared
directly with the synchrotron measurement \cite{Domke1996}. Each
resonant state can be characterized by the corresponding set of the
Fano parameters which are listed in \Tref{Tab1}. The bottom (green)
spectrum corresponds to XUV+IR ionization. The corresponding AIS can
be derived from the red spectrum by absorption or emission of a single
IR photon $\pm\w$ indicated by arrows in \Fref{Fig1}. The parameters of these
two-photon states, also tabulated, have no analogue in synchrotron
measurements. Moreover, these states decay to the two non-resonant
continua. Hence their Fano parameterization differ qualitatively with
the need of introducing a correlation factor $\rho^2$ \ask{\cite{PhysRev.137.A1364}}
\be
\label{Fano}
\sigma(E)=\sigma_0\big[1-\rho^2+\rho^2
\frac{(q+\epsilon)^{2}}{1+\epsilon^{2}}\big] 
\ \ , \ \ 
\epsilon=\frac{E-E_0}{\Gamma/2}
\ .
\ee 
Such an index $\rho^2=1$ in the XUV only spectrum of He.

\begin{table}[b!]
\caption{Fano parameters of the AIS in He and \H. The present
  theoretical values extracted from the XUV-only spectra are compared
  with the He measurement \cite{Domke1996} and an earlier \H
  calculation \cite{Sanchez1997}. In the latter case, the two lowest
  AIS of the $^1\Sigma_u$ symmetry are shown. }
\label{Tab1}
\smallskip

\bt{lccccccccc}
\hline\hline\rule{0mm}{3mm}                            
&\mc{4}{c}{XUV only}& \mc{2}{c}{XUV+IR}\\ 
&\mc{2}{c}{$2sp+$} 
&\mc{2}{c}{$3sp+$} 
&\mc{2}{c}{$2sp+$} \\
              & TDSE   &Ref. & TDSE  &Ref. &$-\w$ & $+\w$\\
\hline\hline\\                            
\mc{2}{l}{He atom}&\cite{Domke1996} && \cite{Domke1996}\\
$q$           & -2.95 &-2.75& -2.46 &-2.5 & -0.93& -1.03\\    
$\Gamma$, meV & 43    & 37  & 8     & 10  & 126 & 137\\
$E_0$, eV     & 60.28 &60.15& 63.70 &63.65& 58.7 &61.8\\
$\rho^2$      &  &  &  &  & 0.53 & 0.59\\\\
\mc{2}{l}{\H molecule}&\cite{Sanchez1997} && \cite{Sanchez1997}\\
$q$           & -0.75 && -0.86 &&-0.47 & \\    
$\Gamma$, meV & 400   &430&137   & 100 & 367 & \\
$E_0$, eV     & 30.3 &30.1& 32.3 &32.3& 29.1 &\\
$\rho^2$      &  &  &  &  &0.73  & \\\\
\et
\end{table}

\begin{figure}[t]
\ig{7cm}{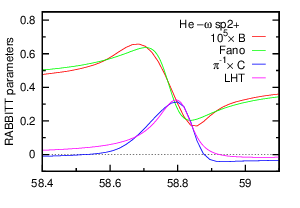}%{Data/Fig2a.eps}

\ig{7cm}{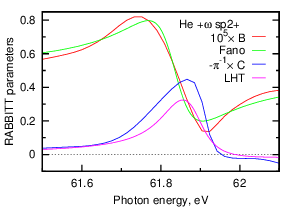}%{Data/Fig2b.eps}

\caption{Resonant $B$ ans $C$ parameters are displayed with the
  corresponding Fano ansatz \eref{Fano} and the LHT transform
  \eref{LHT} for the $\pm\w$ shifted $sp2+$ resonance in He.
}
\label{Fig2}
\end{figure}

\begin{figure}[t]
\ig{7cm}{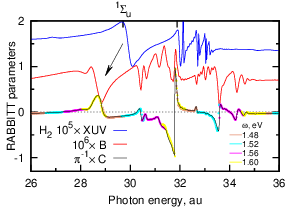}%{Data/Fig3.eps}
\caption{Resonant $B$ ans $C$ parameters in XUV+IR ionization of the
  \H molecule at the equilibrium inter\-nuclear distance of
  $R=1.4$~au. The XUV-only spectrum, also shown, is shifted upwards
  for clarity.  The two lowest AIS of the $^1\Sigma_u$ symmetry are
  marked on the top horizontal axis according to
  \cite{Sanchez1997}. The $C$ parameters extracted from \rain
  calculations at fixed photon energies are marked by rainbow
  colors. Four \rain calculations are needed instead of a single
  \crain one. }
\label{Fig3}
\end{figure}

As is seen in the table, the conventional Fano parameters of the
$ns2+$ and $ns3+$ states are very close between the present
theoretical determination and the experiment \cite{Domke1996}.
Generally, the TDSE should be propagated for the times exceeding the
corresponding lifetime and this would become unpractical for very
long-living higher members of the $sp\,n+$ series, not shown in the
table.  The resonances in the XUV+IR spectrum are noticeably broader
as their seeming lifetime is determined by an overlap with the IR
probe pulse. Outside this overlap, the AIS does not contribute to the
two-photon ionization amplitude \cite{serov2024fano}.

As was shown in \cite{Ji2024}, the knowledge of the Fano parameters
permits to derive the resonant phase and the corresponding
atomic time delay. The amplitude-phase transformation is based
on the Kramers-Kronig relation expressed via a logarithmic Hilbert
transform (LHT). Validity of this transformation was demonstrated in
\cite{Ji2024} in several cases of resonant photo\-ionization induced
by a single XUV photon absorption. Here we apply the same technique to
relate the RABBITT amplitude $B$ and phase $C$ parameters. More
specifically, we convert the XUV+IR Fano parameters from \Tref{Tab1}
into an alternative set $r=\rho^{-2}-1$, $Q=q/(r+1)$ and
$\gamma=\sqrt{r(r+q^2+1)/(r+1)^2}$ to express the phase as prescribed
by of Eq.~(19) of \cite{Ji2024}
\be
\label{LHT}
C = \arg \left[
1+ [Q+i(\gamma-1)](\epsilon+i)^{-1}
\right]
\ee
Both the Fano fit for the $B$ parameter and the LHT result for the $C$
parameter are displayed in \Fref{Fig2} for the $\pm\w$ shifted $2sp+$
resonance.  This figure demonstrates that the amplitude-phase relation
is generally valid in the case of two-photon XUV+IR ionization \ask{when
the resonance is embedded in only one of the two RABBITT arms. This
relation would be difficult to establish if two or more resonances are
intertwined in the neighboring absorption and emission channels.}

The resonant XUV+IR spectrum of the \H molecule at the equilibrium
inter\-nuclear distance $R=1.4$~au is displayed in \Fref{Fig3}. As
compared to the corresponding spectrum of the He atom exhibited as the
green line in \Fref{Fig1} and in more detail in \Fref{Fig2}, the \H
molecule demonstrates a considerably richer spectrum of several
partially overlapping resonant states. The two lowest states of the
$^1\Sigma_u$ symmetry can be resolved sufficiently accurately from the
XUV-only spectrum with the Fano parameters listed in
\Tref{Tab1}. These parameters compare favorably with the literature
values \cite{Sanchez1997}.

\subsection{Below threshold excitations}
\label{Under}

Finally, we apply our \crain technique to look under the threshold and
to resolve a series of single-electron $1s\to np$ exciation\-s in
He. Here we utilize the $u$RABBITT methodology which is based on the
following consideration (see e.g. \cite{kheifets2024resonant}).  When
the laser photon energy is such that
$(2q-1)\w <I_p<2q\w
\ ,
$
the harmonic peak H$_{2q-1}$ submerges below the ionization threshold
while the the adjacent SB$_{2q}$ is still visible in the photoelectron
spectrum. As $\w$ varies, H$_{2q-1}$ crosses a number of discrete
below-threshold excitations and the magnitude and phase parameters of
SB$_{2q}$ change rapidly.  The \crain technique allows to look inside
this near-threshold SB$_{2q}$ without the need for adjusting
$\w$. Moreover, our determination allows to span the whole series of
the below-threshold discrete excitation\-s in a single TDSE run.

\begin{figure}[t]
\ig{8cm}{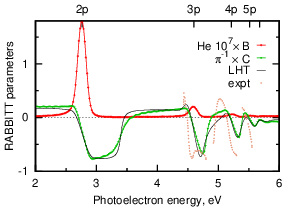}%{Data/Fig4а.eps}
\caption{ RABBITT magnitude $B$ and phase $C$ parameters in He near
  threshold shown as functions of the photo\-electron energy.  The
  data points represent the  measured phase 
  \cite{Neoricic2022}. The thin solid line visualizes the LHT \eref{LHT}.  }
\label{Fig4}
\ms
\end{figure}

This is illustrated in \Fref{Fig4} where we display the $B$ and $C$
parameters in He in the near-threshold region. For the clarity of
presentation, we choose a rather high central photon energy of
$\w=6.1$~eV such that the corresponding photo\-electron spectrum is
stretched sufficiently far away from the threshold.  As the
photo\-electron energy varies, the corresponding position of the
submerged harmonic H$_3$ intersects with the series of discrete
excitation\-s $1s\to np$ whose energies, shifted upwards by $\w$, are
marked on the top horizontal axis of the plot. Each of such crossing
is accompanied by a sharp peak of the magnitude $B$ parameter.  These
peaks are broadened with the spectral width of the APT. At the same
time, the phase $C$ parameter undergoes a series of $\pi$ jumps
dampened by a finite spectral width. This jumps can be reproduced
closely by a numerical LHT
\be
\label{LHT}
C(E) = 
-\frac{1}{\pi} {\mathcal P}\!\!
 \int\limits_{0}^{E_{\max}} \hs{-3mm} dx \
{\ln B(x)\over x-E}
\approx
-\frac{1}{\pi} {\mathcal P}\!\!
 \int\limits_{-\infty}^{\infty}\hs{-2mm} dx \
{\ln B(x)\over x-E}
\ .
\ee
\ask{\Eref{LHT} represents the Kramers-Kronig relation connecting the
imaginary and real parts of the logarithm of a complex function.
In this equation,}  the principal value integral, taken across the photoelectron
spectrum, is extended to infinite limits where the resonant
spectrum is vanishing. The integration in \Eref{LHT} is performed
using the numerical recipe \cite{2020SciPy-NMeth}. \ask{Correction is also
needed for the vanishing amplitude $f(E)$ which makes the logarithmic
derivative $f'(E)/f(E)$ divergent \cite{Burge1974,Burge1976}}
Such LHT calculated $C$ parameter is sufficiently close to the TDSE
calculation. Both are qualitatively similar to the experimental values
\cite{Neoricic2022}.

\section{Conclusion}
\label{Conclusion}

In conclusion, we realize the continuous rainbow \crain based on the
parity  separation of the XUV+IR ionization amplitude. Our
procedure allows to achieve a very fine energy resolution across a
wide spectrum without the need for adjusting the laser photon
frequency.
We demonstrate a great computational efficiency of the 
\crain technique in which the magnitude and phase parameters of the
resonant states can be extracted for the whole series of resonances,
both below and above the ionization threshold. This is a considerable
improvement in comparison with the need for the laser frequency
adjustment in \rain that is required to span several resonant peaks in
the photo\-electron spectrum.

We apply the \crain to resolve the series of AIS in the He atom and
the \H molecule. In addition, we look under the threshold in the
spirit of the $u$RABBITT investigation and resolve tje whole series of
the $1s\to np$ discrete excitation\-s in the He atom. We also apply
the logarithmic Hilbert transform to utilize the Kramers-Kronig
relation and to link the magnitude and phase RABBITT parameters in the
resonant ionization of the He atom. This is the first successful
application of the LHT to two-photon ionization processes.

While the present work demonstrates an application of 
\crain to two-electron targets such as He and \H, more complicated
atomic and molecular systems can be treated in a similar way. This
would only require an expansion of the multi-configuration basis
\eref{multi} built from a greater number of one-electron
orbitals. Only one such orbital needs to be continuous while others
could be built from the discrete ionic states.

Finally, while the parity-based amplitude separation is not feasible
experimentally, the XUV+IR and XUV-only components of the photo\-electron 
wave\-packet can be separated based on their distinct angular distributions. 

\ask{
In the future development, we intend to apply the \crain technique to
the RABBITT process driven by circularly polarized radiation. Such a
circularly polarized RABBITT allows to retrieve the amplitude ratios
and phase difference in the two ionization channels accessible  by the
XUV+IR absorption
\cite{kheifets2024characterization,kheifets2025circularly}. The
under-threshold behavior of the circular RABBITT has never been
investigated before and the present \crain technique will be highly
computationally efficient for this purpose.}

\subsection*{Acknowledgment:}  

We thank Rowan Jesson Kerr for his help with numerical implementation
of the LHT. 
This work was supported by the Discovery Grant DP230101253 from the
Australian Research Council.

\clearpage
%\bibliography{references,Areferences,dreferences,hreferences,kreferences,mreferences,rreferences,Rreferences,treferences,ureferences,Wreferences,wreferences,mypapers,sten,nandi}
%\end{document}

\end{document}